\documentclass[conference]{IEEEtran}
\usepackage{cite}
\ifCLASSINFOpdf
   \usepackage[pdftex]{graphicx}
   \DeclareGraphicsExtensions{.pdf,.jpeg,.png}
\else
   \usepackage[dvips]{graphicx}
   \DeclareGraphicsExtensions{.eps}
\fi
\usepackage{wrapfig}
\usepackage[all]{xy}
\long\def\comment#1{}
\newcommand{\commentout}[1]{}

\newcommand{\secref}[1]{Section~\ref{#1}}
\newcommand{\eqref}[1]{Equation~\ref{#1}}
\newcommand{\figref}[1]{Figure~\ref{#1}}

\newcommand{\denselist}{
      \setlength{\itemsep}{0pt}
      \setlength{\parsep}{1.5pt}
      \setlength{\topsep}{1.5pt}
      \setlength{\parskip}{2pt}
      \setlength{\partopsep}{0pt}
      \setlength{\labelwidth}{1em}
      \setlength{\labelsep}{0.5em} }
\newcommand{\bdesc}{\begin{itemize}\denselist}
\newcommand{\edesc}{\end{itemize}}

\begin{document}
\title{Structure of Heterogeneous Networks}

\author{\IEEEauthorblockN{Rumi Ghosh}
\IEEEauthorblockA{University of Southern California\\
Information Sciences Institute\\
Marina del Rey, California 90292\\
Email: rumi.ghosh@gmail.com}
\and
\IEEEauthorblockN{Kristina Lerman}
\IEEEauthorblockA{University of Southern California\\
Information Sciences Institute\\
Marina del Rey, California 90292\\
Email: lerman@isi.edu}
}

\maketitle

\begin{abstract}
Heterogeneous networks play a key role in the evolution of communities and the decisions individuals make.  These networks link different types of entities, for example, people and the events they attend. Network analysis algorithms  usually project such networks unto simple graphs composed of entities of a single type. In the process, they conflate relations between entities of different types and loose important structural information.
We develop a mathematical framework that can be used to compactly represent and analyze heterogeneous networks that combine multiple entity and link types.
We generalize Bonacich centrality, which measures connectivity between nodes by the number of paths between them, to heterogeneous networks and use this measure to study network structure. Specifically, we extend the popular modularity-maximization method for community detection to use this centrality metric. We also rank nodes based on their connectivity to other nodes. One advantage of this centrality metric is that it has a tunable parameter we can use to set the length scale of interactions. By studying how rankings change with this parameter allows us to identify important nodes in the network.
We apply the proposed method to analyze the structure of several heterogeneous networks. We show that exploiting additional sources of evidence corresponding to links between, as well as among, different entity types yields new insights into network structure.

\end{abstract}

%
\IEEEpeerreviewmaketitle

\section{Introduction}
Heterogeneous networks play a key role in information dissemination, evolution of communities,  and the decisions individuals make.  While traditional network analysis algorithms can efficiently find structure even in large data sets, they usually work on homogeneous data, i.e., networks composed of entities of a single type, for example, a social network where individuals are nodes and an edge between nodes corresponds to a (possibly directed) friendship relationship. Such networks can be represented as unipartite graphs. Many online networks, however, mix entities of different types. On the popular photo-sharing site Flickr, for example, users can post images, tag them with descriptive keywords, join special-interest photography groups, befriend other users, mark images of others users as their favorite, and so on. We can represent Flickr as a \emph{heterogeneous} network composed of several entity types: \emph{users}, \emph{images}, \emph{groups}, and \emph{tags}, with connections between the entities representing different types of relations. A link between users denotes a friendship; a link between a user and a group denotes user's membership in the group; a link between an image and tags represents the keywords used to annotate that image, and so on. In order to extract useful knowledge from this data, we need to look at the network in its entirety.

Heterogeneous networks are sometimes represented as multi-partite (bipartite, etc.) graphs in which vertices are partitioned into disjoint sets corresponding to different entities, with edges connecting vertices from different sets. While some approaches consider the bipartite graph as the whole~\cite{Freeman}, others usually project the network onto a unipartite graph for matrix algebraic analysis~\cite{Breiger}. Thus, the  user-group network is reduced to a graph containing users only, with links between users denoting membership in the same photo group. Such projections, however, lose important information about network structure~\cite{Guimera07}. Recently, \cite{Barber07,Guimera07} extended the modularity-based approach~\cite{GirvanNewman04} to find structure in bipartite graphs. They showed that compared to analyzing a projected graph, taking into account links between different entity types leads to better community structure. However, bipartite graphs do not allow for edges between vertices of the same type; thus, methods based on this representation cannot exploit all the information available in a heterogeneous network. In the user-group network, for example, information from friendship links between users could augment information from group membership, leading to a better understanding of the network.

This paper makes two contributions. First, in Section~\ref{sec:representation}, we present a mathematical framework for compactly representing a heterogeneous network that combines entities and links of different types. We represent such a network as a multi-layer graph, where each layer contains vertices (entities) of a unique type, with edges linking vertices across different layers, as well as within a single layer. Thus, the user-group network is a 2-layer graph, with intra-layer edges in the user layer giving user-user (friendship) relations, and inter-layer edges giving the user-group (membership) relations.
Using this mathematical representation, we develop algorithms to study network structure.
We use Bonacich centrality~\cite{Bonacich87} as the basis for network analysis, specifically for identifying communities and important nodes in a network. This centrality metric, defined in Section~\ref{sec:structure}, gives the number of paths of any length linking two nodes. It contains a tunable parameter that allows us to set the length scale of the interactions. As the second contribution of the paper, we extend the modularity-based community detection algorithm to find groups of nodes that are more connected, in Bonacich centrality sense, to each other than to outside nodes. We also use Bonacich centrality to rank individual nodes in the network.
Finally, in Section~\ref{sec:results}, we apply this framework to study the structure of real-world heterogeneous networks. We analyze two benchmark networks studied in literature, as well as a network extracted from the social photosharing site Flickr. We show that exploiting information contained in links \emph{between}, and \emph{among}, different entity types leads to new insights into network structure.

\section{N-Mode Matrix Representation}
\label{sec:representation}
We compactly represent a heterogeneous network as a layered graph, in which entities belonging to different classes are partitioned into separate layers, with \emph{intra-layer} and \emph{inter-layer} edges representing links between entities. Consider a network with two entity classes $X$ ($|X|=n$) and $Y$ ($|Y|=m$). For concreteness, suppose the data represents a scientific papers dataset with authors $X$ and papers $Y$, and that in addition to the usual authorship relations, we managed to collect additional data about friendships, acknowledgements and citations. This data can be represented as a graph with two layers, with vertices of type $X$ (authors) in one layer, and vertices of type $Y$ (papers) in the other layer.
An $(m+n) \times (m+n)$ adjacency matrix captures the intra- and inter-layer relations between different vertices:
\[ A=\left[ \begin{array}{cc}
XX_{m \times m} & XY_{m \times n} \\
YX_{n \times m} & YY_{n \times n}
\end{array} \right]
\]
Here $A_{ij}=XX_{ij}$ gives the binary relation of the ordered pair $(x_{i},x_{j})$, e.g., a friendship between authors $i$ and $j$; $A_{i,j+m}=XY_{ij}$ gives the binary relation of the ordered pair $(x_{i},y_{j})$, e.g., if author $i$ wrote paper $j$; $A_{i+m, j}=YX_{ij}$ gives the binary relation of the ordered pair $(y_{i},x_{j})$, e.g., if paper $i$ acknowledges author $j$; $A_{i+m,j+m}=YY_{ij}$ gives the binary relation of the ordered pair $(y_{i},y_{j})$, e.g., whether paper $i$ cites paper $j$.
We call this data structure a \emph{2-mode matrix}.
This representation is similar to one used by Tong et al.~\cite{Tong08} to represent bipartite graphs,
except since bipartite graphs only describe the inter-layer, and not the intra-layer, relations, the diagonal submatrices $XX$ and $YY$ are zero.
\comment{
\[ A=\left(
\begin{array}{cc}
0_{m \times m} & XY_{m \times n} \\
YX_{n \times m} & 0_{n \times n}
\end{array} \right)
\].
}

We can easily generalize the above formulation to $N$-mode matrices, which represent graphs having $N$ distinct types of nodes or being composed of entities belonging to $N$ distinct classes.
The adjacency matrix in this case represents $N^2$ distinct types of binary relations.
Now that we have a mathematical representation of heterogeneous networks, we are ready to explore their structure.

\section{Network Centrality and Structure}
\label{sec:structure}

Centrality measures the degree to which network structure determines importance of a node in a network. Social network researchers have proposed several different measures of centrality~\cite{Katz,Freeman,Bonacich87} to explain the influence or status of individual actors within a social network. Katz~\cite{Katz}, for example, recognized that an individual actor's centrality depends not only on how many others she is connected to (her degree), but also on the centrality of the players she is connected to. Katz score measures status of an actor by the total number of paths linking it to other nodes in the network, exponentially weighted by the length of the path~\cite{Katz}.
Bonacich~\cite{Bonacich87} generalized this idea by introducing a new measure of centrality, $C(\alpha,\beta)$, parameterized by $\alpha$ and $\beta$. Bonacich centrality (\emph{b-centrality}) measures the expected number of transmissions directly or indirectly caused by a node. Like Katz score, b-centrality is given by the total number of attenuated paths emanating from a node, but now the attenuation factors along direct links, $\beta$, and indirect links, $\alpha$, in a path can be different\footnote{For some types of networks, e.g., commodity exchange networks, Bonacich allows $\alpha<0$. In communication and information networks we are considering, $\alpha>0$. Also, in this paper we reverse Bonacich's notation and take $\beta$ as direct attenuation and $\alpha$ as indirect attenuation factors.}
\begin{eqnarray}
\label{eq:inf}
C(\alpha,\beta) &=& (\beta A +\beta \alpha A \cdot A + \cdots + \beta \alpha^n  A^{n+1} \cdots) \nonumber \\
& =& \beta A {( I-\alpha A)}^{-1}\,.
\end{eqnarray}
\noindent
This equation holds while  $\alpha < 1/\lambda$, where $\lambda$ is the largest characteristic root of  $A$~\cite{Ferrar}. For $\alpha=\beta$, this measure reduces to the Katz status score~\cite{Katz}. B-centrality can be easily generalized to heterogeneous networks, with matrix $A$ corresponding to the $N$-mode matrix representing the network.
We can use b-centrality to study the structure of a heterogeneous network. We extend the popular modularity-based community detection method to utilize b-centrality. In addition, we show that b-centrality can identify influential nodes within the network, as well as nodes that bridge different communities. One advantage of using b-centrality is that we can vary parameter $\alpha$ to set the length scale of the interactions.
$\alpha$ is the probability of transmitting a message or influence along an indirect edge in a path emanating from a vertex. The expected length of a path, the radius of centrality, is $(1-\alpha)^{-1}$.
For $\alpha=0$, b-centrality takes into account direct edges only. Many network analysis algorithms use such local structures, e.g., the degree of the vertex, as the metric in their analysis. As $\alpha$ increases, b-centrality becomes a more global measure, taking into account ever larger network components. This tunable parameter turns b-centrality into a powerful tool for investigating network structure.
In real-world networks, we can estimate the value of $\alpha$ along a particular link of a network by measuring the probability that a node transmits a message received from a distant node along this link.   In most situations, however, this information is not readily available. Although we may not know its exact value, studying how network properties change with $\alpha$ gives us valuable insight into network structure.

\subsection{Community Detection}
\label{sec:modularity}
\vspace{-0.06 in}Girvan \& Newman~\cite{GirvanNewman04} proposed \emph{modularity} as a metric for evaluating community structure of a network.
The modularity-optimization class of community detection algorithms~\cite{Newman104,Newman204,Newman206} find a network division that maximizes the modularity $Q$, given by $Q=$(connectivity within community)-(expected connectivity), where connectivity is density of edges.
We extend this approach and use b-centrality as the measure of network connectivity~\cite{Ghosh08}.
Therefore, in the best division of the network, nodes have more paths connecting them to nodes within their community than to outside nodes.
We generalize modularity $Q$ as
\begin{equation}
\label{eq: mod2}
Q(\alpha)=\sum_{ij} {[C_{ij} - \bar{C}_{ij}]\delta(s_i, s_j)}
\end{equation}
where $C_{ij}$ is given by Eq.~\ref{eq:inf}, $\bar{C}$ is the expected b-centrality, and $s_i$ is the index of the community $i$ belongs to, with $\delta(s_i, s_j) = 1$ if $s_i =s_j$; otherwise, $\delta(s_i, s_j)=0$. We round the values of $C_{ij}$ to the nearest integer. Since $\beta$ factors out of modularity, we consider dependence on $\alpha$ only.
To compute the expected centrality, we consider a graph, referred to as the null model, which has the same number of vertices and edges as the original graph, but in which the edges are placed at random. To make the derivation below more intuitive, instead of b-centrality we talk of the number of paths.
When all the vertices are placed in a single group, then  axiomatically, $Q=0$. Therefore $ \sum_{ij}[C_{ij} - \bar{C}_{ij}] =0$, and we set
$W = \sum_{ij} \bar{C}_{ij}=\sum_{ij} C_{ij}.$
Therefore, according to the argument above, the total number of paths between vertices in the null model $\sum_{ij} \bar{C}_{ij}$ is equal to the total number of paths in the original graph, $\sum_{ij} C_{ij}$.  We further restrict the choice of null model to one where the expected number of paths reaching vertex $j$, $W_j^{in}$, is equal to the actual number of paths reaching the corresponding vertex in the original graph.
$
W_{j}^{in} = \sum_{i} \bar{C}_{ij} = \sum_{i} C_{ij}\,$.
Similarly, we also assume that in the null model, the expected number of paths originating at vertex $i$, $W_{i}^{out}$, is equal to the actual number of paths originating at the corresponding vertex in the original graph
$
W_{i}^{out} = \sum_{j} \bar{C}_{ij} = \sum_{j} C_{ij}\,.
$
Next, we reduce the original graph $G$ to a new graph $G^{\prime}$ that has the same number of vertices as $G$  and total number of edges  $W$, such that each edge has weight 1 and the number of edges between nodes $i$ and $j$ in $G^{\prime}$ is  $C_{ij}$. Now the expected number of paths between  $i$ and $j$  in graph $G$ could be taken as the expected number of the edges between vertices $i$ and $j$ in graph $G^{\prime}$, and the actual number of paths between vertices $i$ and $j$  in graph $G$ can be taken as the actual number of edges  between vertex $i$ and vertex $j$ in graph $G^{\prime}$. The equivalent random graph $G''$ is used to find the \emph{expected}  number of edges  from vertex $i$ to vertex $j$. In this graph  the edges are placed in random subject to constraints:
\bdesc
\item The total number of edges in $G''$ is $W$.
\item The out-degree of vertex $i$ in $G''$ = out-degree of vertex $i$ in $G^{\prime} = W_{i}^{out}$.
\item The in-degree of a vertex $j$ in graph $G''$ =in-degree of vertex $j$ in graph $G^{\prime} =W_{j}^{in}$.
\edesc

Thus in $G''$ the  probability that an edge will emanate from a particular vertex  depends only on the out-degree of that vertex; the probability that an edge is incident on a particular vertex depends only on the  in-degree of that vertex; and the probabilities of the two vertices being the two ends of a single edge are independent of each other. In this case, the probability that an edge exists from $i$ to $j$ is given by $C$(\emph{emanates from i}) $\cdot $ $C$(\emph{incident on j})=$(W_{i}^{out}/W)(W_{j}^{in}/W)$.
Since the total number of edges is $W$ in $G''$, therefore the expected number of edges between $i$ and $j$ is $W \cdot (W_{i}^{out}/W)(W_{j}^{in}/W)=\bar{C}_{ij}$, the expected the expected b-centrality in $G$.
Once we compute modularity $Q(\alpha)$ for the N-mode matrix representing the  network, we have to select an algorithm to divide the network into communities that optimize $Q(\alpha)$. Brandes et al.~\cite{Brandes} have shown that the decision version of modularity maximization is NP-complete. Like others~\cite{Newman206,Leicht}, we use the leading eigenvector method to obtain an approximate solution. In this method, vertices are assigned to either of two groups based on a single eigenvector  corresponding to the largest  positive eigenvalue of the modularity matrix (spectral optimization of modularity).

As the network grows, matrix $A$ may become quite large, making computation of inverse in Eq.~\eqref{eq:inf} expensive. We use an approximation method, along the lines of \cite{Foster01}, that keeps the first three terms in Eq.~\eqref{eq:inf} only.

\subsection{Node Ranking}
\label{sec:ranking}
\vspace{-0.06 in}Social scientists have long believed that structure of the network can affect an individual's productivity and success~\cite{Simmel,Burt} and predict new links~\cite{Liben07} (or ties). Much of the analysis done by social scientists considered local structure, i.e., the nature of an individual's ties~\cite{Simmel, Granovetter, Burt}.

By focusing on local structure, the traditional microscopic theories fail to capture the global, macroscopic structure of the network.
This structure is better captured by metrics that take into account $paths$  and not merely $links$ or ties between nodes.
Several different centrality metrics take paths into account to identify nodes that are `close' in some sense to other nodes in the network, and are therefore, more important.
Betweenness centrality~\cite{Freeman} calculates node's score as the ratio of the number of shortest paths via the given node to the number of shortest paths in the network.
As described above, Katz score~\cite{Katz} of node $i$ is the sum over all paths from $i$, exponentially weighted by the length of the path.
PageRank~\cite{PageRank}, roughly, gives the probability that a random walk initiated at node $i$ will reach $j$.
Liben-Nowell and Kleinberg~\cite{Liben07} evaluated performance of the different scoring mechanisms on the link prediction task and showed that Katz score is one of the most effective measures for this task.

We follow Bonacich~\cite{Bonacich87} and use b-centrality $C(\alpha,\beta)$ as the measure of proximity between nodes in a network. As mentioned above, this metric is a generalization of the Katz score, and enables us to identify important nodes  in the network. A node could have a high b-centrality if it is connected to many nodes within its community --- these are community $leaders$. A node could also have a high b-centrality if it is connected to nodes in different communities. While they may be peripheral to any given community, these nodes play an important $bridging$ role in the network: they mediate communication between communities.
We can identify such nodes, because their b-centrality increases as $\alpha$, the weight of distant links, grows. Other centrality metrics do not distinguish between leaders and bridges.

\section{Empirical Results}
\label{sec:results}
We apply the formalism developed above to study the structure of two heterogeneous networks that have been studied in literature, the College Football~\cite{GirvanNewman02} and the Southern Women~\cite{Freeman02} datasets datasets, as well as the user-group network extracted from the social photosharing site Flickr.

We adopt normalized mutual information, $MI$, as the metric for evaluating the quality of discovered communities~\cite{Danon05,Barber07}. Suppose our method found a community division $X$, whereas the actual community division of the network is $Y$. The probability that a node is assigned to group $x$ by the algorithm, whereas it actually belongs to group $y$ is $P(x,y)=N_{xy}/n$, where $N_{xy}$ is the number of nodes that were assigned to $x$ that belong to group $y$, and $n$ is the total number of nodes. Following Barber~\cite{Barber07}, we express normalized mutual information as
$$
 MI(X,Y)=\frac{2I(X,Y)}{H(X)+H(Y)}, \nonumber
$$
where standard mutual information and entropy are defined as $I(X,Y)=\sum_{x,y}{P(X,Y) \log{\frac{P(X,Y)}{P(X)P(Y)}}}$, $H(X)=\sum_x{P(X)\log{P(X)}}$, and  $H(Y)=\sum_y{P(Y)\log{PY)}}$. When $MI=1$, the discovered groups are the actual communities in the network. When $MI=0$, the discovered groups are independent of the actual communities.

\subsection{Southern Women}
\label{sec:southern}
\begin{figure}[tbh]
\begin{center}
\includegraphics[height= 2 in]{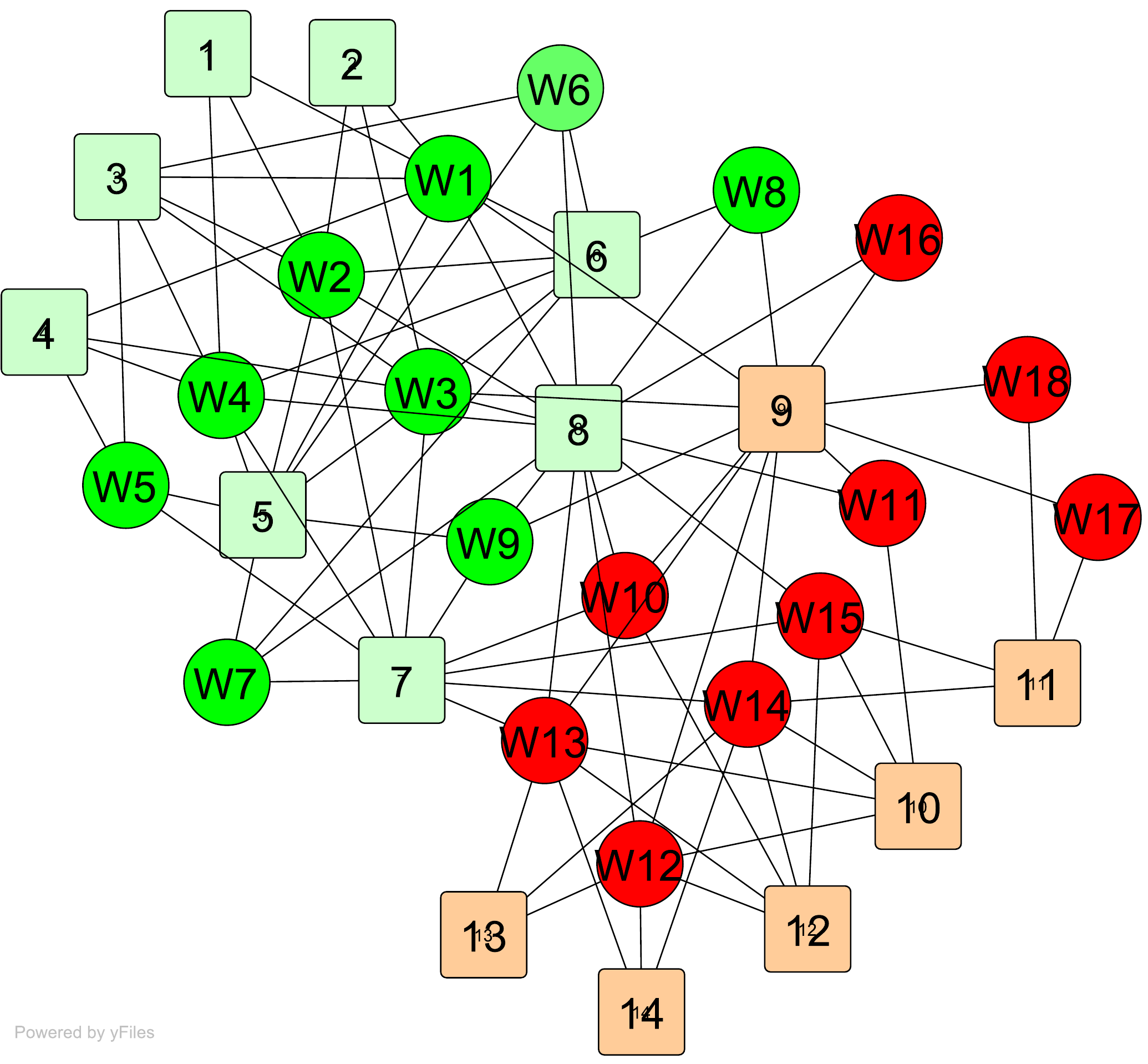}
\end{center}
\caption{Bipartite graph representing the Southern Women dataset.  Circles represent women and squares the events they attended. Nodes in red and pink belong to one group, while those in green and light green to the other. }
\label{fig:dataset}
\end{figure}
\vspace{-0.06in}
The Southern Women dataset  comes from a comparative study of social class by Davis et al.~\cite{DGG}.
The researchers collected  systematic data on the social activities of 18 women over a nine month period.
During this time, various subsets of women met in a series of 14 informal events, as shown in \figref{fig:dataset}.
Many researchers have subsequently tried to predict the social classes and the structure of groups in this dataset~\cite{Freeman94}.
Freeman~\cite{Freeman02}  reviewed 21 such studies and performed a meta-analysis of
predictions to find the groups. He did a canonical analysis of symmetry and dynamic-paired-comparison scaling to
find women's positions (rank) within the group.  We take Freeman's meta-analysis as ground truth for our study.

\subsubsection{Communities}
We set $\alpha \leq 0.16$, the reciprocal of the largest eigenvalue of the 2-mode matrix. For all values of $\alpha$  women w1 -- w9 were assigned to Group1, and w10 -- w18 to Group2, the same results as the ground truth found by Freeman's meta-analysis~\cite{Freeman02}.  Events 1 to 8 were assigned to Group1, and events 9 to 14 Group2. The mutual information metric was $MI=1$. Only 6 of the 21 algorithms in Freeman meta-analysis replicated the ground truth.

Alternatively, the bipartite women-events network can be projected onto a unipartite graph of women only, where a link between women exists if they attended an event together.
The community detection algorithm put w2, w4--w7 in one group and the rest of the women in the other, resulting in $MI=0.38$.
As Guimera~\cite{Guimera07} noted, such projections loose information contained in the bipartite graph. A projection that results in a weighted adjacency matrix, i.e., where $A_{ij}$ shows the number of events women $i$ and  $j$ attended together, preserves this information. For $\alpha \leq 0.01$, the reciprocal of the largest eigenvalue of the unipartite matrix, the community detection results agreed with the ground truth, resulting in $MI=1$.

\subsubsection{Rankings}
When we ranked the women we obtained interesting insights into the structure of groups. Table~\ref{tbl:group1} shows the rankings of women within each group, with 1 as the highest rank. When only the direct links are considered ($\alpha=0$), then in Group1, w1, w2, w3, and w4 form the core, w5, w6, w7 the primary and w8, w9 the secondary members (with w9 ranking higher than w8), as predicted originally  by Davis et al. \cite{DGG}.\footnote{Core, primary and secondary was used in \protect\cite{DGG} to assign the status of the women within the group.} However, when we increase $\alpha$, a crisper ordering emerges. Among the core members, w3 takes the leadership position, followed by w1, w4 and w2. More interestingly, as the strength of indirect links grows, the importance of peripheral members changes. For $\alpha=0.1$, w9  is ranked higher than w5, and she keeps moving up in rank with increasing $\alpha$. Woman w9 is a peripheral member of Group1, and is also connected to Group2. In fact the original study assigns w9 as a secondary member of both groups, because she was ``claimed'' by both groups~\cite{DGG}. Such peripheral members loosely connected to both communities act as bridges between communities and are responsible for the spread of information from one community to another~\cite{Granovetter,Burt}. Our analysis easily identifies these important people.

\begin{table*}
\begin{tabular}{|c|c|c|c|c|c|c|c|c|c|c|c|}
    \hline
ID    &    Meta1 &    Meta2   &   $\alpha=0$  &   $\alpha=0.02$   &   $\alpha=0.04$   &   $\alpha=0.06$   &    $\alpha=0.08$  &   $\alpha=0.1$    &   $\alpha=0.12$   &   $\alpha=0.14$   &   $\alpha=0.16$   \\
\hline
w1   &   1   &   1   &   2.5 &   2   &   2   &   2   &   2   &   2   &   2   &   2   &   2   \\
w2   &   2   &   2   &   2.5 &   4   &   4   &   4   &   4   &   4   &   4   &   4   &   4   \\
w3   &   3   &   3   &   2.5 &   1   &   1   &   1   &   1   &   1   &   1   &   1   &   1   \\
w4   &   4   &   4   &   2.5 &   3   &   3   &   3   &   3   &   3   &   3   &   3   &   3   \\
w5   &   5.5 &   5   &   6   &   7   &   7   &   7   &   7   &   8   &   8   &   8   &   8   \\
w6   &   5.5 &   6   &   6   &   5.5 &   6   &   6   &   6   &   6   &   6   &   7   &   7   \\
w7   &   7   &   7   &   6   &   5.5 &   5   &   5   &   5   &   5   &   5   &   5   &   6   \\
w8   &   9   &   9   &   9   &   9   &   9   &   9   &   9   &   9   &   9   &   9   &   9   \\
w9   &   8   &   8   &   8   &   8   &   8   &   8   &   8   &   7   &   7   &   6   &   5   \\
\hline
w10 &   6   &   6   &   7   &   6   &   6   &   6  &    6   &   6   &   6   &   5   &   4   \\
w11  &   5   &   5   &   4.5 &   4.5 &   4.5 &   5   &   5   &   5   &   5   &   6   &   6   \\
w12  &   3   &   3   &   2.5 &   3   &   3   &   3   &   3   &   3   &   3   &   3   &   2   \\
w13  &   1   &   1   &   2.5 &   2   &   2   &   2   &   2   &   2   &   1   &   1   &   1   \\
w14  &   2   &   2   &   1   &   1   &   1   &   1   &   1   &   1   &   2   &   2   &   3   \\
w15  &   4   &   4   &   4.5 &   4.5 &   4.5 &   4   &   4   &   4   &   4   &   4   &   5   \\
w16  &   9   &   9   &   9   &   9   &   9   &   9   &   9   &   9   &   7   &   7   &   7   \\
w17  &   7.5 &   7.5 &   7   &   7.5 &   7.5 &   7.5 &   7.5 &   7.5 &   8.5 &   8.5 &   8.5 \\
w18  &   7.5 &   7.5 &   7   &   7.5 &   7.5 &   7.5 &   7.5 &   7.5 &   8.5 &   8.5 &   8.5 \\
    \hline
\end{tabular}
  \caption{Rankings of women within their groups in the Southern Women dataset. Meta1 refers to Canonical Analysis, and Meta2 to Paired Comparison. The remaining columns give rankings for different values of  $\alpha$.}\label{tbl:group1}
\end{table*}
\normalsize

In Group2, w14 emerges as the leader for $\alpha=0$, followed by w13 and w12. Women w15 and w11, who are given the same rank, come next, followed by w10, w17, and w18, all with the same rank. Woman 16 has the lowest rank. On increasing $\alpha$, w16 becomes increasingly more important, surpassing w17 and w18, i.e., she emerges as the ``bridge''. Woman w16 is connected to only one node in Group2, yet  she becomes more important than w17 and w18 (both connected to two nodes in the group), because she is also connected to Group1. Much the same way, the gradual increase in the ranking of w10  with $\alpha$ can be attributed to her connection to the other group. When $\alpha=0.12$, there is a change in ranking of w13 and w14, with w13 taking the leadership position. This is the ranking obtained in the meta-analysis, with only w16 ranking above w17 and w18 by our method as compared to the meta-analysis. At $\alpha=0.14$ w10 ranks above  w11. It is, in fact, the two women (w10 and w16), who do not conform to the ranking of the meta analysis, who act as bridges between communities.

We applied the same analysis to the unipartite graph, which is the projection of the bipartite data unto a graph of women only.
The rankings of Group1 women were somewhat different from the results of the meta-analysis. Whereas the meta-analysis assigned the highest rank to w1, in our analysis w3 had that position. The rankings were mostly independent of $\alpha$, with only the rankings of w6 and w7 changing as $\alpha$ increased.
The rankings of Group2 women were also almost independent of $\alpha$. At $\alpha=0$, w11 and w15 had the same rank, but when $\alpha$ grew, w15 was higher ranked. The rankings are similar to the ground truth, with the only difference being w16, who is  placed above w17 and w18 by our algorithm, but below w17 and w18 in the rankings obtained from the meta-analysis.
In summary, although b-centrality-based rankings produced by the bipartite and unipartite methods were similar to the ground truth, bipartite method allowed us to identify ``bridges'' who facilitate communication between different communities.

\subsection{College Football}
\label{sec:football}
\vspace{-0.06in}
The US College football dataset~\cite{GirvanNewman02} represents the schedule of Division 1 games for the 2001 college football season. The teams are divided into \emph{conferences}  containing 8 to 12 teams each.
Games are more frequent between members of the same conference. Inter-conference games, however, are not uniformly distributed, with teams that are geographically closer likely to play more games with one another than teams separated by geographic distances.

\comment{
\begin{figure}
  \begin{tabular}{c}
      \includegraphics[width=2.5in]{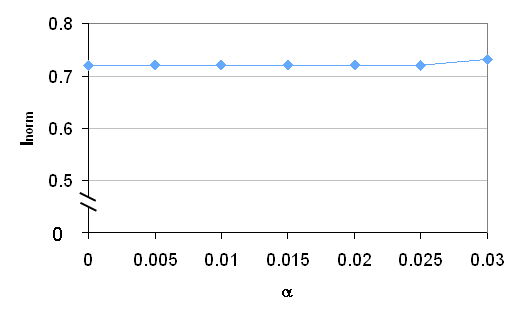} \\
      (a)\\
        \includegraphics[width=2.5in]{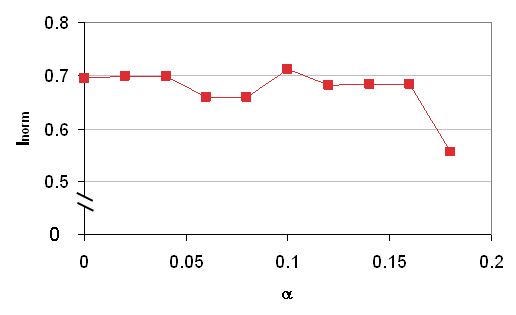} \\
       (b)
  \end{tabular}
  \caption{Mutual information measure of groups discovered in the College Football data represented as (a) a heterogeneous and (b) a unipartite graph}\label{fig:football}
\end{figure}
}

\comment{
\begin{table}
  \centering
\begin{tabular}{|r|r||r|r|}
  \hline
  \multicolumn{2}{c}{unimodal} & \multicolumn{2}{c}{bimodal}\\ \hline
  $\alpha$ & $MI$ & $\alpha$ & $MI$ \\ \hline
  0   & 0.6952 & 0  & 0.7192\\
0.02 &  0.6987& 0.005 &  0.7192 \\
0.04  &  0.6987& 0.01  &  0.7192 \\
0.06  &  0.6583& 0.015 &  0.7192\\
0.08  &  0.6583& 0.02  &  0.7192\\
0.1 &0.7112& 0.025 &  0.7192 \\
0.12 &   0.6816& 0.03  &  0.7316 \\
0.14 &   0.6838& & \\
0.16 &   0.6838& & \\
0.18 &   0.5568& & \\
\hline
\end{tabular}
  \caption{Normalized mutual information measure of communities discovered in the College Football data}\label{tbl:football}
\end{table}
}

\begin{table}
  \centering
\begin{tabular}{|r|r||r|r||r|r||r|r|}
  \hline
    \multicolumn{4}{c}{College Football} & \multicolumn{4}{c}{Flickr}\\ \hline
  \multicolumn{2}{c}{ (a) unimodal} & \multicolumn{2}{c}{(b) bimodal} &   \multicolumn{2}{c}{(c) unimodal} & \multicolumn{2}{c}{(d) bimodal}\\ \hline
  $\alpha$ & $MI$ & $\alpha$ & $MI$ &   $\alpha$ & $MI$ & $\alpha$ & $MI$\\ \hline
  0   & 0.695 & 0  & 0.719 & 0 & 0.288 & 0 & 0.1682 \\
0.02 &  0.699& 0.005 &  0.719 & 0.001 & 0.286 & 0.002 & 0.16\\
0.04  &  0.699& 0.01  &  0.719 & 0.002 & 0.288 & 0.004 & 0.148 \\
0.06  &  0.658& 0.015 &  0.719 & 0.003 & 0.289 & 0.006 & 0.083 \\
0.08  &  0.658& 0.02  &  0.719 & 0.004 & 0.289 & 0.008 & 0.110 \\
0.1 &0.711& 0.025   &  0.719 & 0.005 & 0.290 & &\\
0.12 &   0.682& 0.03  &  0.732 & 0.006 & 0.292 & & \\
0.14 &   0.684&     & & 0.007 & 0.293 & & \\
0.16 &   0.684&     & & 0.008 & 0.294 & & \\
0.18 &   0.557&     & & 0.009 & 0.296 & & \\
    &         &     & & 0.01 & 0.271 & & \\
\hline
\end{tabular}
  \caption{Normalized mutual information measure of communities discovered in the networks. Unimodal refers to networks containing nodes of a single type, while bimodal refers to heterogeneous networks with two types of nodes.}\label{tbl:results}
\end{table}
We represent the College Football dataset as a 2-mode matrix.
The games between teams give the team-to-team relations, while the conferences to which they belong give the team-to-conference relations. Unlike the Southern Women dataset, a purely bipartite network, this dataset contains both  the relations among teams and between teams and conferences.

We used modularity-based approach to find communities for different values of $\alpha$, $0 \le \alpha \le 0.03$.
We find eight groups independent of $\alpha$; however, setting $\alpha$ to its maximum value leads to purer groups. 
Table~\ref{tbl:results}(b) shows the mutual information-based measure of the quality of the groups discovered in this network.
The groups for the most part follow conference membership. Cases where deviations from conference membership occur have natural interpretations, such as geographic proximity of teams. In some cases, the groupings reflect past associations, with a team being assigned to a group with other teams from its former conference, rather than the new conference it belongs to. More interestingly, we find that deviations from conference membership predict future developments, specifically, teams switching conference membership after 2001.
For example, New Mexico State (SunBelt Conference) was grouped with Western Athletic (WAC) by our algorithm. It joined WAC in 2005. Texas Christian (Conference USA) was also grouped with WAC. It was part of WAC but joined Conference USA in 2001. Central Florida (Independent) was grouped with Mid-American conference by the algorithm, and joined it in 2002. Notre Dame (Independent) was grouped with Big 10 conference, and as of 2008, is in talks of joining it.
Alternatively, we can represent the College Football dataset as a unipartite graph, where the vertices are teams and  edges represent regular season game between the teams~\cite{GirvanNewman02}.
Table~\ref{tbl:results}(a) shows the mutual information-based measure of the quality of the discovered groups vs $\alpha$ in this network. Note that maximum value of $\alpha$ is bigger for this network. The communities are less pure than those discovered using the 2-mode matrix. Overall, the heterogeneous method gives a crisper division. When it does put conferences together, these assignments make sense for geographic or historic reasons, and as we showed above, sometimes anticipate future developments.

\subsection{Flickr Social Network}
\vspace{-0.06 in}We also ran our algorithm on the heterogeneous social network data collected from Flickr, a social photosharing site that allows users to upload images, post them to special interest photo groups, and to join social networks by adding other users as friends or contacts.
Since the actual social network on Flickr is rather vast, we sampled it by identifying users who were broadly interested in one of three topics~\cite{Lerman07flickrsearch}: child and family \emph{portraiture}, \emph{nature} photography and \emph{technology}. For each topic, we used the Flickr API to perform a tag search using a keyword relevant to that topic, to retrieve 500 `most interesting' images. We then extracted the names of users who submitted these images to Flickr and added them to our data set. The keywords used for image search were (a) \textsf{newborn} for the \emph{portraiture} topic, (b) \textsf{tiger} and \textsf{beetle} for the \emph{nature} topic, and (c) \textsf{apple} for the \emph{technology} topic. Each keyword is ambiguous. \textsf{Tiger}, for example, could mean a wild animal, but also a flower (Tiger lily), Mac operating system (OS X Tiger), or a famous golfer (Tiger Woods), while \textsf{beetle} could describe a bug or a car.

From the set of users in each topic, we identified four (eight for \emph{nature}) who were interested in each topic. We examined each user's profile to confirm that the user was indeed interested in that topic. Specifically, we looked at group membership and user's most common tags. Thus, groups such as ``Big Cats'', ``Zoo'', ``The Wildlife Photography'', etc.  pointed to user's interest in the \emph{nature} topic.
We used the Flickr API to retrieve the contacts of these users, as well as their contacts' contacts. We labeled contacts by the topic of the seed user.
Although we did not verify that all the labeled users were indeed interested in the topic, we use these \emph{soft labels} to evaluate the discovered communities.


\comment{
\begin{table}
  \centering
\begin{tabular}{|r|r||r|r|}
  \hline
  \multicolumn{2}{c}{unimodal} & \multicolumn{2}{c}{bimodal}\\ \hline
  $\alpha$ & $MI$ & $\alpha$ & $MI$ \\ \hline
0 & 0.2877 & 0 & 0.1682 \\
0.001 & 0.2864 & 0.002 & 0.16 \\
0.002 & 0.288 & 0.004 & 0.1484 \\
0.003 & 0.2887 & 0.006 & 0.0834 \\
0.004 & 0.2894 & 0.008 & 0.1104 \\
0.005 & 0.2904 & & \\
0.006 & 0.2918 & & \\
0.007 & 0.2932 & & \\
0.008 & 0.2942 & & \\
0.009 & 0.2958 & & \\
0.01 & 0.2714 & & \\
\hline
\end{tabular}
  \caption{Normalized mutual information measure of communities discovered in Flickr data. }\label{tbl:flickr}
\end{table}
}

\subsubsection{Communities} 
Once we retrieved the social networks of a target set of users, we reduced it to an undirected network containing mutual contacts only. In other words, every link in the network between two nodes, say $A$ and $B$, implies that $A$ lists $B$ as contact and \emph{vice versa}. This resulted in a network of $5747$ users. Of these, $1620$ users were labeled \emph{technology}, $1337$ and $2790$ users were labeled \emph{portraiture} and \emph{wildlife} respectively.
The normalized mutual information for community division of this network is shown in Table~\ref{tbl:results}(c)
We took the soft labels corresponding to topic of photography interest as the true community division of the network. As $\alpha$ increases up to its maximum value, the groups become purer, and the mutual information increases. Except for the maximum value of $\alpha$, there were three groups. Group1 was composed mainly of  \emph{technology} users, Group2 mainly of \emph{wildlife} users. Users interested in \emph{portraiture} emerged as a distinct group, Group3, whose size was largely independent of $\alpha$. The fourth group found at $\alpha=0$ was a mixture of all topics, and at the maximum value of $\alpha$ only two groups were found.

Next, we augmented the mutual contacts data with information about user membership in public groups on Flickr. We used the Flickr API to retrieve the public groups to which the users in our dataset belonged, 51,000 groups in total. We considered a user to be active in a group if among the most recent 100 photos she uploaded, more than 10  were posted to that group.  We were able to extract the active groups for 3625 of the 5747 users, and they belonged to $10,463$  active groups.


We represented this data as a 2-mode matrix of users and groups, where a relation between users specified whether they were each other's mutual contacts, and a relation between a user and a group specified whether the user was an active member of this group.
The 2-mode matrix proved too large for eigenvector decomposition on the computing resources available to us. Instead, we used the Lanczos algorithm\cite{Lanczos} to efficiently compute the leading eigenvalue of the 2-mode matrix and then used the eigenvector corresponding to this eigenvalue to optimize modularity.
Table~\ref{tbl:results}(d) evaluates the quality of the community division of the heterogeneous network. While the normalized mutual information metric is worse than for the mutual contacts network, looking closer at the results suggests that the heterogeneous network has a somewhat different structure. In the mutual contacts network, the \emph{portraiture} group emerged as a distinct group, probably because its members are tightly interconnected. In the heterogeneous user-group network, the \emph{nature} group emerges as a distinct group. Although members of this group seem to be less well-connected as contacts, they appear to be active in similar groups. Our method takes user-group relations into account and is able to identify these users to yield new insights into the structure of the Flickr community.

\begin{figure*}
  \begin{center}
  \begin{tabular}{cc}
      \includegraphics[width=2.3in]{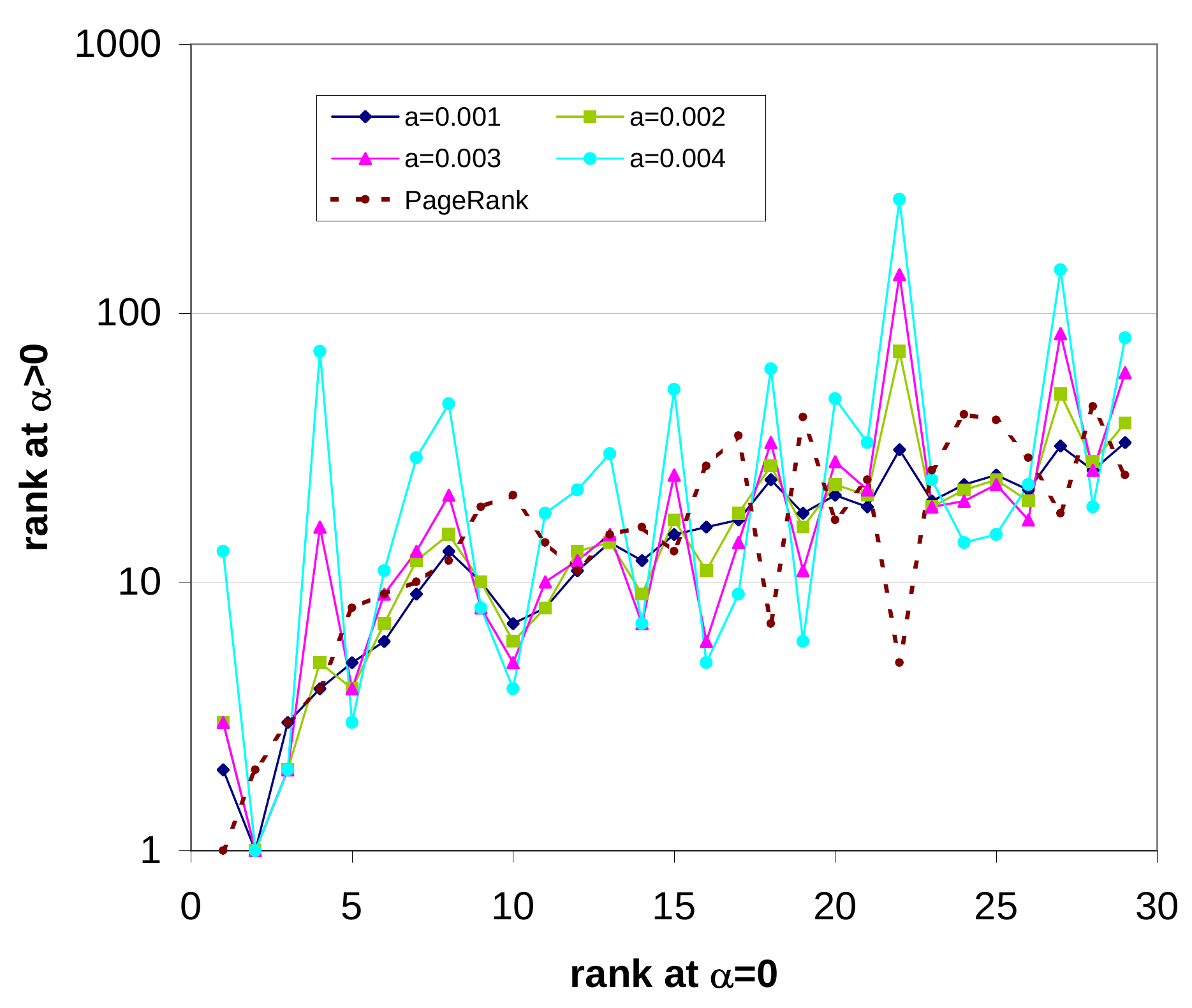} &
        \includegraphics[width=2.3in]{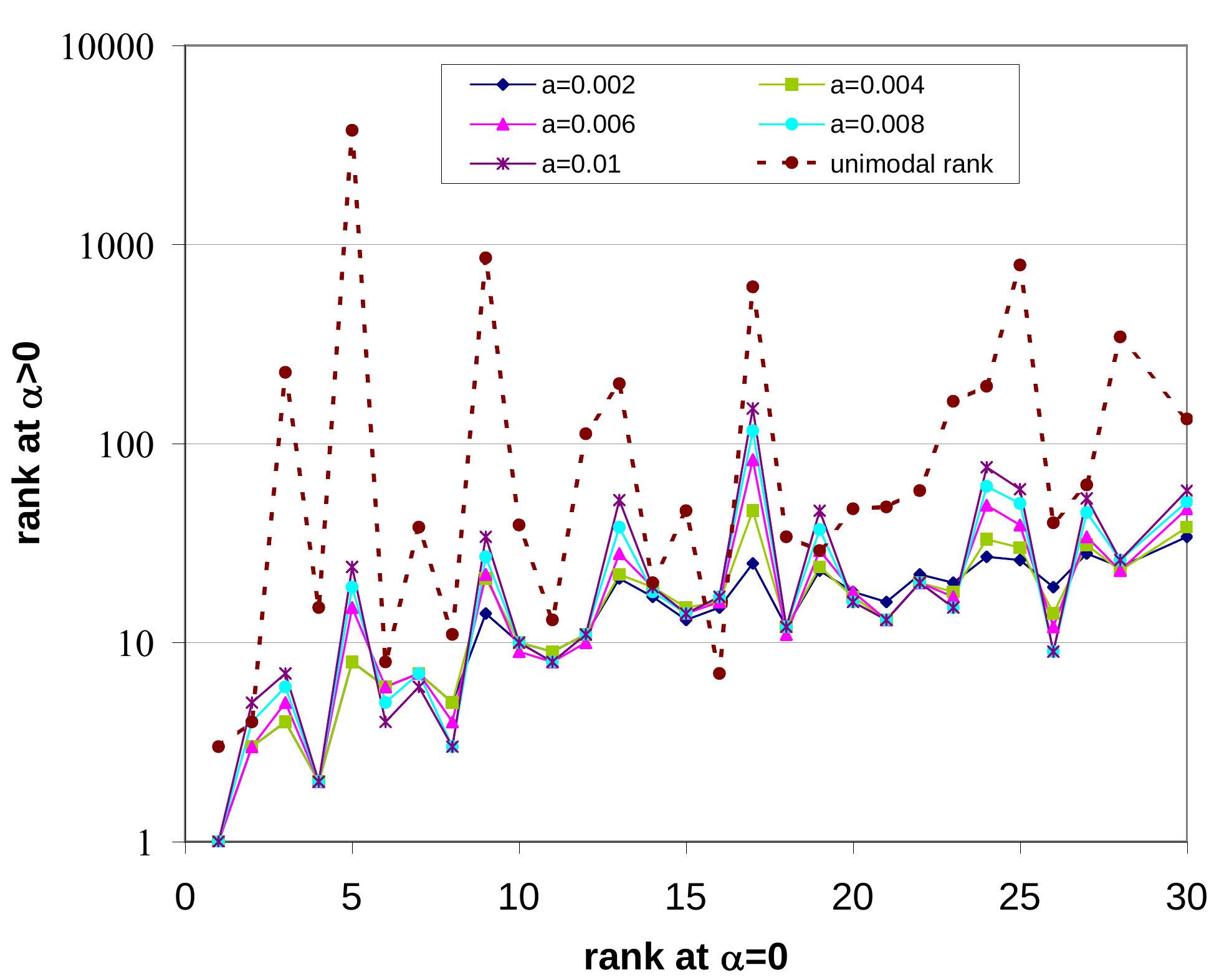} \\
       (a) & (b)
  \end{tabular}
  \end{center}
  \caption{Rankings of select Flickr users in the (a) mutual contacts and (b) user-group networks}\label{fig:rankings}
\end{figure*}

\subsubsection{Rankings} 
We ranked users according to their b-centrality. Unfortunately, there is no independent analysis of the rankings of users, so we do not have a gold standard to evaluate the results of our algorithm. Figure~\ref{fig:rankings} shows how the rankings of users relative to their ranking at $\alpha=0$ change with increasing $\alpha$ in the (a) mutual contacts and (b) user-group networks. We claim that nodes whose rank improves with $\alpha$  (5, 10, 14, 16, 19) are the bridging nodes. Though we have no way to confirm it, it appears that these users appeal to others outside their community. Other nodes (4, 8, 18) see their rank worsen with $\alpha$. These are nodes peripheral to the $portraiture$ group that dominates the rankings. The mutual contacts-based rankings correlate somewhat with PageRank-based rankings.

Rankings of the user-group network (Fig.~\ref{fig:rankings}(b)) produce similar trends, though there are only 9 users who were in the top-ranked set shown in Fig.~\ref{fig:rankings}(a). The new top-ranked users are mostly from the $nature$ group. Although the data is difficult to evaluate, taking user-group relations into account appears to emphasize the importance of the $nature$ group.

\section{Related Work}
\label{sec:related}
Liben-Nowell and Kleinberg~\cite{Liben07} have shown that Katz measure is the most effective measure for the link prediction task, better than hitting time, PageRank~\cite{PageRank} and its variants.
Unlike the Katz score, Bonacich centrality~\cite{Bonacich87}, remained relatively unknown in the computer science community. It parametrizes the Katz score with $\alpha$, a parameter that gives the weight of distant links, and also sets the scale of the centrality measure. We showed the benefit of using this parameter in the analysis of network structure.

There has been some work in motif-based communities in complex networks \cite{Arenas} which like our work extends traditional notion of modularity introduced by Girvan and Newman~\cite{GirvanNewman02}. The underlying motivation for motif-based community detection is that  ``the high density of edges within a community determines correlations between nodes going beyond nearest-neighbours,'' which is also our motivation for applying centrality-based modularity to community detection.
Though the motivation of this method is to determine the correlations between nodes beyond nearest neighbors, yet it does impose a limit on the proximity of neighbors to be taken into consideration dependent on the size of the motifs. The method we propose, on the other hand, imposes no such limit on proximity. On the contrary, it considers the correlation between nodes in a more global sense. The measure of global correlation evaluated using the b-centrality metric  would  be equal to the weighted average of correlations  when motifs of different sizes are taken. B-centrality enables us to calculate this complex term quickly and efficiently.

Resolution limit is one of the main limitations of the original modularity detection approach\cite{fortunato07_2}. It can account for the comment by Leskovec et al.~\cite{Leskovec08www}  that they ``observe tight but almost trivial communities at very small scales, the best possible communities gradually `blend in'  with rest of the network  and thus become less `community-like'.'' However, that study is based on the hypothesis that communities have ``more and/or better-connected `internal edges' connecting members of the set than `cut edges' connecting to the rest of the world.'' Hence, like most graph partitioning and modularity-based approaches to community detection, their process depends on the local property of connectivity of nodes to neighbors via edges and is not dependent on the structure of the network on the whole. Therefore, it  does not take into account connectivity in a more global sense, as given by centrality metrics. In their paper on motif-based community detection, Arenas et al.\cite{Arenas} state that  the  extended quality functions for-motif based modularity  also obey the principle of the resolution limit. But this limit is now motif-dependent and then several resolution of substructures can be achieved by changing the motif. However, it would be difficult to verify  which resolution of substructures is closest to natural communities. In b-centrality-based modularity, on the other hand, the resolution limit depends on the centrality radius, given by the attenuation factor $\alpha$. Smaller $\alpha$ lead to smaller radii, and, therefore, to division of the network into a larger number of communities~\cite{Ghosh08}.

There have been two recent works that extend modularity-based approach to bipartite networks~\cite{Guimera07,Barber07}. Both of these methods express modularity in terms of edges; therefore, their formulation of modularity maximization suffers from the same problem of localization as the original formulation by Newman for unipartite graphs, and are unable to determine correlation between nodes beyond nearest neighbors.
We, on the other hand, can vary parameter $\alpha$ to take nodes beyond nearest neighbors into account.
Barber et al. ~\cite{Barber07} argue that in the representation of modularity of used by Guimera et al.~\cite{Guimera07} identifies modules in only one part of the network at a time. They, on the other hand, classify nodes in both partitions simultaneously and customize spectral methods to bipartite graphs. The customization is based on the identification of the asymmetric  submatrix of the full bipartite modularity matrix. This asymmetric submatrix not only represents the bipartite nature of the graph, but also enables them to customize bipartite modularity-maximization method by using singular value decomposition and recursive identification of bipartite modules. However since this algorithm explicitly takes advantage of the bipartite nature of the graph, it cannot be used for graphs containing intra-layer edges along with inter-layer edges. For example, in the case of Flickr, bipartite representation may capture only user-group relations, but not information about user-user or group-group relations. Hence our method is more appropriate to capturing the complete information encoded within a social network.

\section{Conclusions}
In this paper, we introduced a compact data structure, the N-mode matrix, to represent different classes of entities and relations present in a heterogeneous network. We used Bonacich centrality to study the structure of such networks, specifically, identify communities and important nodes in the network. We extended the modularity optimization-based class of algorithms to use b-centrality, rather than edges, as a measure of network connectivity. We applied this approach to  benchmark networks studied in literature and found that it results in network division in close agreement with the ground truth. In addition, it  gave useful insights into the structure of the graph and information about the changes that happen in the future, but  were not known at the time when data was collected. We also used b-centrality to rank nodes in a network.  By studying changes in rankings that occur when the indirect attenuation factor $\alpha$ changes, we were able to identify leaders and `bridging' nodes that facilitate communication between different communities.
The results of the community-finding algorithm applied to Flickr network were mixed. One possibility is that since the number of groups and group membership is much larger than the number of users, group information completely masks user-user information. We may want to differentially weigh relations to balance transmission of influence along different channels. To do this, we break the 2-mode matrix into diagonal (intra-layer) and off-diagonal (inter-layer) components: $A = D_1 A_1+D_2 A_2, $ where
\begin{eqnarray}
\nonumber
A_1=\left[
\begin{array}{cc}
XX_{m \times m} & 0_{m \times n} \\
0_{n \times m} & YY_{n \times n}
\end{array} \right]
&
A_2=\left[
\begin{array}{cc}
0_{m \times m} & XY_{m \times n} \\
YX_{n \times m} & 0_{n \times n}
\end{array} \right],
\end{eqnarray}
and weights are given by matrices
\begin{eqnarray}
\nonumber
D_1=\left[
\begin{array}{cc}
D^{\gamma}_{m \times m} & 0_{m \times n} \\
0_{n \times m} & D^{\delta}_{n \times n}
\end{array} \right]
&
D_2=\left[
\begin{array}{cc}
D^{\lambda}_{m \times m} & 0_{m \times n} \\
0_{n \times m} & D^{\mu}_{n \times n}
\end{array} \right],
\end{eqnarray}
with each $D^{\gamma}$, $\ldots$, $D^{\mu}$ a diagonal matrix with $D^{\gamma}_{ii}=\gamma$, etc.
We plan to study this balancing scheme on real-world networks.

\section*{Acknowledgment}
This work is based in part on research supported by the National
Science Foundation under awards BCS-0527725 and CMMI-0753124.



\end{document}